\documentclass[prl,aps,amsmath,amssymb,superscriptaddress,twocolumn]{revtex4}
\usepackage{graphicx}
\usepackage{dcolumn}
\usepackage{bm}
\usepackage{subfigure}
\usepackage{setspace}
\usepackage{feynmp}
\usepackage{psfrag}
\DeclareMathSizes{10}{9}{6}{6}

\begin{document}

\title{Finite phase coherence time of a bosonic quantum field at the Boltzmann equilibrium}
\author{Dr.~Alexej Schelle}

\affiliation{Senior Lecturer @ IU Internationale Hochschule, Juri-Gagarin-Ring 152, D-99084 Erfurt \\ email : alexej.schelle@gmail.com}

\begin{abstract}

A quantitative quantum field approach with non-local order parameters is presented for a very weakly interacting, dilute Bose gas. 
Within the presented model, which assumes the constraint of particle number conservation at constant average energy in the canonical ensemble, it is shown that both coherent oscillations, as well as decay times of quantum coherence for the forward and backward propagating components of the quantum field created by the atomic cloud of a very weakly interacting Bose-Einstein condensate, are defined by a unique time variable. 
Within the present numerical theory, a quantitative estimate for the unit time scale for time propagation of a particle in a very weakly interacting Bose gas is derived from the coherence time of the wave field and it is illustrated that this time scale defines a unit time for transitions between different realizations of the 
Boltzmann equilibrium as defined by the maximum entropy from the vanishing of the chemical potential.

\end{abstract}

\maketitle

\begin{description}

\item[Purpose]: arXiv version of the manuscript

\end{description}

\section{I. Introduction}

As a fundamental property of quantum mechanics, coherence demonstrates the wave nature of quantum particles by the ability of the particles to interfere when their wavelength becomes of the order of their average distance below the critical temperature \cite{ref-1}. 
In classical physics, when a particle is moved from point A to a second point B, the particle has a well-defined position and velocity at each instant of time during the cycle from A to B, and indeed this time is well-defined by the path length between the two positions and the average velocity of the traveling particle \cite{ref-2}. Quantum mechanically, however, the dynamics of particles are more complicated, because of uncertainty and localization of position and velocity for particles in external confinements, thus in the setup as described above, the quantum particle may only be mainly located at two different positions and therefore exhibit no probability to be located in the space between the two points A and B. 
As a consequence of quantum effects from quantization of the corresponding single-particle wave functions, if the particle tunnels from point A to point B, there remains a time interval during which it is quantum mechanically not known whether the particle is located at position A or B, so that the quantum state of the particle turns into a coherent superposition of a state with localization in point A and a state with localization in point B, with a mathematically defined phase relation between these two states which allows to quantify the oscillations of the transition probability for the particle to move from point A to B, and vice versa. 
Apart from spatial localization and coherence as described above, quantum states of non-local type (non-local order parameters) can obey phase coherence, i.e. well-defined relations of the relative or partial phases between the different quantum states. 
The time scale during which such representing states are quantum mechanically correlated and the particle's wave function can interfere and coherently oscillate is called coherence time \cite{ref-3}.

Calculations and analyses of coherence times typically rely on the knowledge of the spectrum of the given quantum system. 
In a Bose-Einstein condensate below the critical temperature, most of the particles in the Bose gas search to populate the ground state of the system, so that a macroscopic occupation of the ground state mode with a typically centered spatial distribution occurs below the critical temperature for Bose-Einstein condensation with the characteristics of a coherent matter wave \cite{ref-4}. 
In the condensed state, all the particles' wave functions typically tend to constructively interfere at the center of the external confinement. 
This happens because each of the particles can in principle tunnel from the ground state to a (multimode) excited state of the single-particle spectrum, because of the large energy uncertainty below the critical temperature. As a consequence of such single-particle coherence from tunneling as described above, each particle is thus in general in a superposition of a state with the localization characteristics of the single-particle ground state, and one or multiple of the excited single-particle states - with interfering forward and backward propagating components of the wave fields. 
The coherence achieved this way is a precursor for Bose-Einstein condensation, where the coherence time is required to be sufficiently large to allow the different single-particle wave functions to interfere constructively and coherently - i. e. in particular larger than the oscillation time of particles among the different single-particle quantum states a particle in the Bose gas may coherently populate. 

Within the framework of quantum mechanics, quantum processes obey time-reversal symmetry due to well-defined phase relations, but only on the time scale of coherence, or more precisely, as long as the quantum system is not coupled to an external thermal environment that can quantify the forward propagation of time. 
It is the oscillation time of forward and backward propagating partial components of the wave fields that defines the time scale for which the two disjoint counterpropagating components interfere at multiples of the oscillation time in complex number 
representation of the spatially averaged order parameter \cite{ref-5, ref-6}.  
As shown in the sequel of this paper, since the oscillation time in complex plane effectively tends to zero on average (standing wave field), it is mainly the coherence time of the quantum field that defines a natural scale for the forward propagation of time.
Thus, phase coherence between partial matter waves defines a minimal unit of time in the present many-body quantum system which adjusts its direction of time due to the constructive interference and its scale by the destructive interference of different partial matter waves. 

So far, no quantitative estimates for the phase correlation time between counter-propagating partial waves of a quantum particle have been derived from first principles in physical units within standard field theories for Bose-Einstein condensation which make use of particle number conservation in the canonical ensemble. 
In the following analysis, the coherence time is therefore estimated for a quantum particle in a nearly ideal Bose-Einstein condensate confined in a harmonic trap, within a number-conserving quantum field theory for Bose-Einstein condensation \cite{ref-7, ref-8}, which assumes sufficiently dilute atomic gases. 
The subsequent theory's quantum field and its coherence time can be understood as the spatially averaged field with temporal coherence obtained e.g. by the measurement of reflected laser light from the atomic sample. 

To connect the mathematical framework of Bose-Einstein condensates at thermal equilibrium to the physics of coherence and oscillation times of a single particle in a Bose gas at finite temperature, the concept of imaginary time turns out to be a suitable mathematical tool to describe both the quantitative scaling of coherence times as well as their uncertainty and broken gauge symmetry aspects induced by many-particle fluctuations below the critical temperature \cite{ref-9}. 
Whereas the definition of (purely) imaginary time often lacks a clear physical interpretation in other frameworks, such as the calculation of the Hawking radiation in black holes \cite{ref-10}, or the definition and calculation of a Eukledian time in special relativity \cite{ref-11}, in the present framework, the formal assumption of a complex time variable can be interpreted as the time scale during which a quantum particle in the gas relaxes towards an equilibrium quantum state imposing the assumption of a Boltzmann equilibrium (i. e. coupling elementary quantum states to the external environment). 

In comparison to (purely) imaginary time that defines the decay characteristics in the vertical direction of the complex plane, from the conservation of particle number and average energy in the framework of the canonical ensemble, the particles in the gas simultaneously exhibit both finite coherence times as well as oscillations between different regions A and B in space (e. g. spatial domains of different single-particle wave functions) below the critical temperature also at the Boltzmann equilibrium, an effect that can be energetically understood as the coupling of symmetric and asymmetric parts of the underlying complex fugacity spectrum. 
Thus, the foundation of coherent particle coupling among different condensate and non-condensate atom number states in a Bose-Einstein condensate as described in the present model is due to a process related to elementary time evolution which defines a finite scale for unit time that is larger or equal to the complex time variable with a forward propagating direction as defined by the Boltzmann equlibrium \cite{ref-5}. 
This in particular defines a (lower bound of) the unit time scale for the forward propagation of time and equivalently the relaxation time for quantum coherence at the interaction of a particle with its (thermal) environment. 
Therefore, each of the particles is coherently distributed among different single-particle quantum states, which leads to a coherent field distribution with different possible condensate and non-condensate number states at a constant total particle number in the Bose gas, which finally defines a macroscopic ground state population from coherent wave interference at a stable Boltzmann equilibrium of maximum entropy below the critical temperature. 

\section{II. Theory}

Within a non-local quantum field theoretical framework, a Bose-Einstein condensate at finite temperature can be described quantitatively by the following field ansatz,

\begin{equation}
\psi = \psi_{0} + \psi_{\perp} = \sum_{\bf{k}}c_{\bf{k}}{\rm e}^{\frac{-i\mu_{\bf{k}}t}{\hbar}},
\label{bec_field}
\end{equation}\\ 
where $\psi_0$ represents the spatially integrated condensate field, and $\psi_\perp$ the non-condensate field in the framework of solid numbers in the standard complex plane \cite{ref-9}. 
In the absence of external perturbations (below the critical temperature), and assuming isotropic relative spatial phase dependence, the total field $\psi = 0$ vanishes so that the total symmetry of the system remains preserved. 
The two condensate and non-condensate field components can be decomposed into uncorrelated single field components. We assume a spatially uniform distribution of the relative phase between the different single-particle wave functions. 
In contrast, as will be shown in the sequel of the present numerical analysis, below the critical temperature, where different atom number states get effectively coherently coupled because of finite coherence times, the resulting quantum states are of the form $\vert \mu \rangle = \sum_{\bf{k}} c_{\bf{k}}\vert \lbrace N_{\bf{k}}\rbrace\rangle$ which leads to the fact that $\psi_0 = -\psi_\perp \ne 0$  - called spontaneously broken gauge symmetry \cite{ref-12}.   

\begin{figure}[t]
\begin{center}
\includegraphics[width = 4.25cm, height = 4.25cm, angle=0.0]{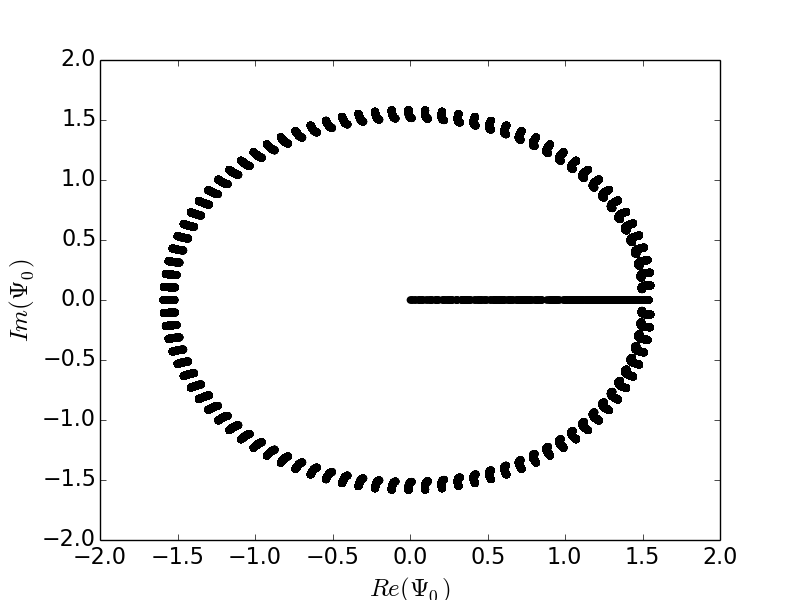} 
\includegraphics[width = 4.25cm, height = 4.25cm, angle=0.0]{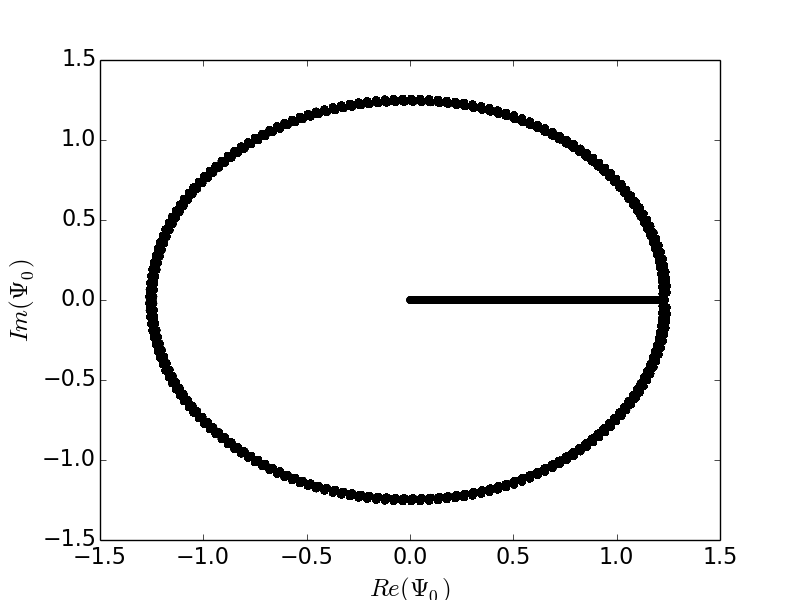} \\
\includegraphics[width = 4.25cm, height = 4.25cm, angle=0.0]{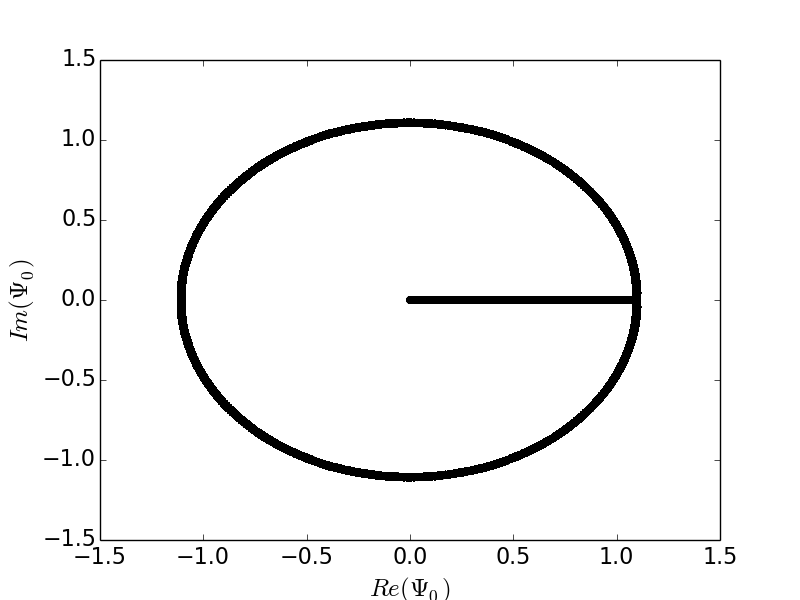} 
\includegraphics[width = 4.25cm, height = 4.25cm, angle=0.0]{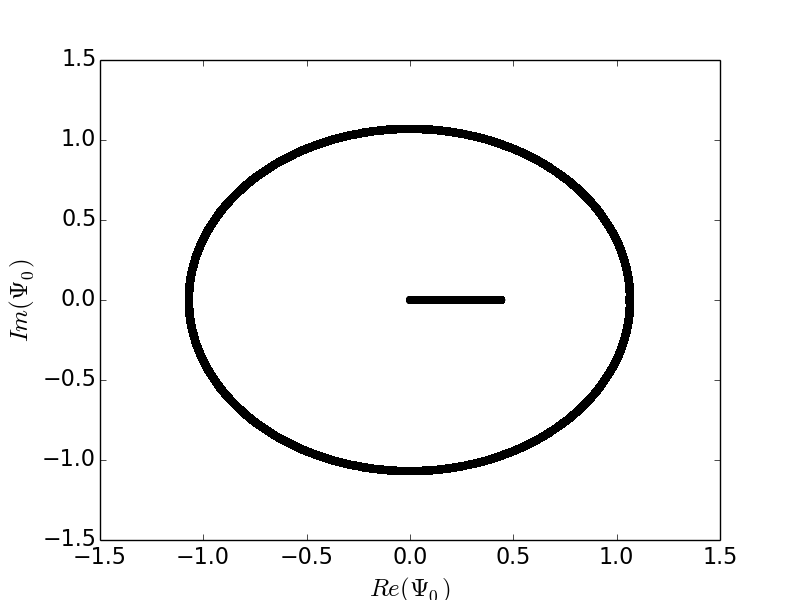} \\
\caption{(Color online) Figures show the spectrum of fugacity values for a Bose-Einstein condensate with $N=1000$ particles in a harmonic trap with trap frequencies $\omega_x,\omega_y = 2\pi\times42.0$ Hz, and $\omega_z = 2\pi\times120.0$ Hz
at different temperatures, $T=5.0 {\rm ~nK}$ (upper left), $T=10.0 {\rm ~nK}$ (upper right), $T=22.5 {\rm ~nK}$ (lower left) and $T=35.0 {\rm ~nK}$ (lower right). 
Below the critical temperature $T_c = 26.9$ nK, the Boltzmann equilibrium state (Bose-Einstein condensation) corresponds to ${\rm Re}(z)\rightarrow1$ ($t\rightarrow0$). 
In this limit, unconnected parts of the spectrum are strongly coupled. 
Whereas the fugacity ring spectrum is gapless below the critical temperature, it obeys a gap and tends to get continuously close 
above $T_c$ ($T=35.0 {\rm~nK}$) with a large range of possible and coupled equilibrium states, obeying the same absolute value of the fugacity tends to one.}
\label{figone}
\end{center}
\end{figure} 

\begin{figure}[b]
\begin{center}
\includegraphics[width=6.5cm, height = 4.0cm, angle=0.0]{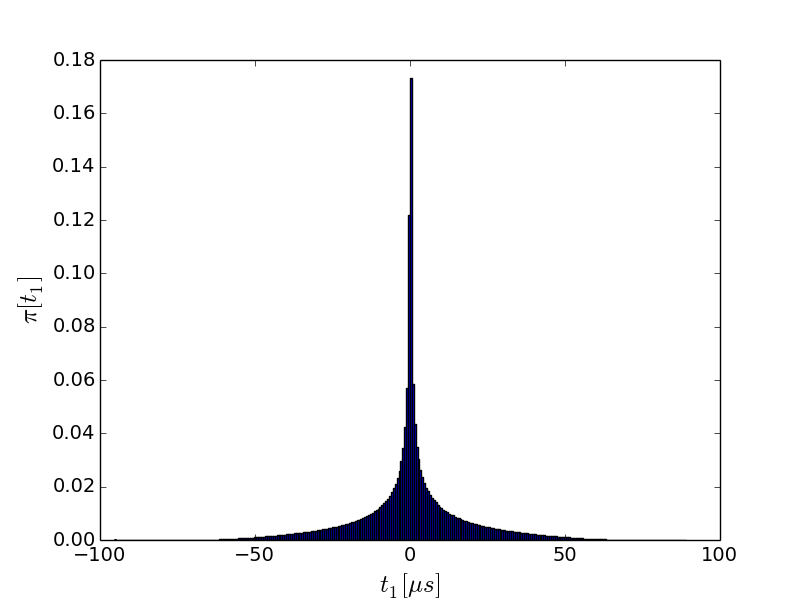} 
\includegraphics[width=6.5cm, height = 4.0cm, angle=0.0]{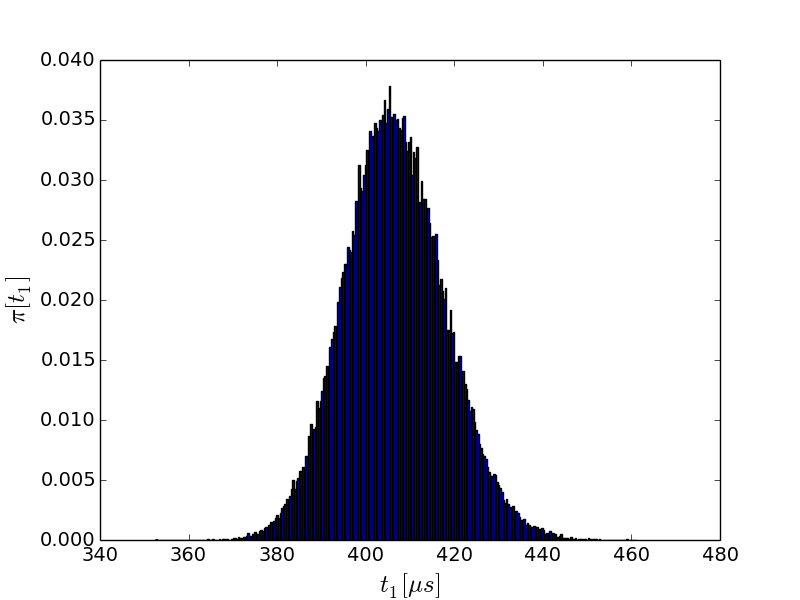} 
\caption{(Color online) Figure shows $10^5$ realizations of the imaginary parts (upper panel - labeled $t_1 = \Gamma^{-1}_1$) and real parts (lower figure - labeled $t_2 = \Gamma^{-1}_2$) 
of complex phase as defined in Eq. (\ref{coherence_time_2}) from random Markov sampling for a non-isotropic trap geometry $\omega_x,\omega_y = 2\pi\times42.0$ Hz and $\omega_z = 120.0$ Hz 
at temperature $T=5.0$ nK. 
The observation that $\Gamma_2t$ is non-zero at thermal equilibrium indicates the coupling of the equilibrium state to quantum states with $\Gamma_2t\ne0$ (non-classical correlations).}.
 \label{figtwo}
\end{center}
\end{figure}

To this end, the concept of imaginary time provides a solid starting point for the analysis of coherence time and its uncertainty - since it defines a relation between the time evolution of a single particle in the Bose gas and the equilibration of the many-particle system at finite temperature $T$, i.e. the interaction of the quantum particle with the surrounding quantum gas. 
It is the statistics of the present quantum field model which then describes the collective dynamics of the quantum field for a Bose-Einstein condensate through relating the parameters of single particles to the many-particle characteristics of the Bose gas. 
Whereas the concept of imaginary time is rather a pure mathematical convention in many quantum theoretical applications, in the current framework, the underlying mathematical relation (Wick rotation) 

\begin{equation}
\beta \hbar = it
\label{coherence_time}
\end{equation}\\
defines a relation between the absolute temperature and the coherent dynamics of a quantum particle in the presence of the $N-1$ other particles in the Bose gas by its absolute value $\vert t\vert=\beta\hbar$. 
The time parameter $t$ can be interpreted as the absolute value of time for forward time evolution on the resolvable scale of unit time in the equilibrated quantum gas, i. e. the time scale of a stable Boltzmann equilibrium before 
approaching the next many-particle configuration that corresponds to the equilibrium of maximum entropy. 
In the present theory, it is the lowest single-particle energy of a quantum particle in the external confinement that defines the uncertainty scale of energy and time, since all values lower than the single-particle ground state energy are effectively zero, because of the finite energy uncertainty.

As a formal ansatz, in Eq. (\ref{coherence_time}), one may use that $\mu t$ is not a real, but a complex number, that is $\mu t\in\mathbb{C}$. Hence, in the framework of this quantum field theory as presented, the real part $\Gamma_1t$ describes the coherent phase evolution of the quantum particle at equilibrium (inverse oscillation frequency in the stationary state), corresponding to the ring of the fugacity spectrum in Fig. \ref{figone}, whereas $\Gamma_2 t$, the imaginary part (of $\mu t$) describes decay processes of the particle in the atomic cloud at or close to equilibrium (compare linear part of the fugacity spectrum in Fig. \ref{figone}), that is
 
\begin{equation}
\frac{\mu t}{\hbar\overline{\omega}} = \Gamma_1 t + i \Gamma_2 t ,
\label{coherence_time_2}
\end{equation}\\
where $\overline{\omega} = 0.5*(\omega_x\omega_y\omega_z)^{1/3}$. 

\begin{figure}[t]
\begin{center}
\includegraphics[width = 7.0cm, height = 4.0cm, angle=0.0]{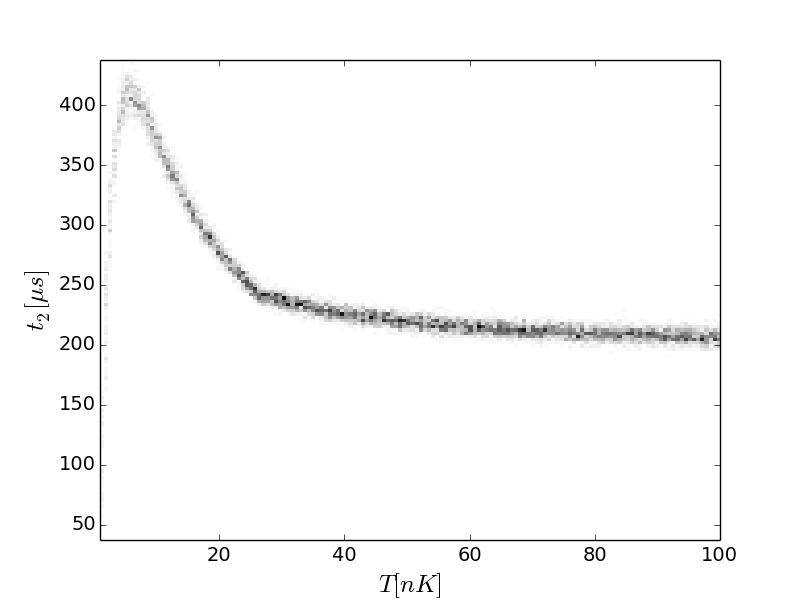}
\caption{(Color online) Figure shows the coherence phase $t_2 = \Gamma^{-1}_2$ as a function of temperature $T$ for a particle in the trap geometry $\omega_x,\omega_y = 2\pi\times42.0$ Hz and $\omega_z = 120.0$ Hz. 
For intermediate values of temperature, the coherence time is around $(300 - 450)~\mu$s. 
Above the critical temperature, it approaches a value of around $200~\mu$s.}
\label{figthree}
\end{center}
\end{figure}

Above the critical temperature, the real part of the fugacity spectrum shows a gap between the ring of constant absolute time, indicating that the symmetry of the quantum field can in principle not be broken without externally induced quantum fluctuations of energy and the corresponding entropy. Decreasing temperature by external cooling to decrease the gap between the outer ring and the inner linear (symmetry breaking) part of the fugacity spectrum finally leads to a gapless fugacity spectrum. 
However, due to the uncertainty of complex time, there is also a finite coupling probability between symmetric and asymmetric parts of the fugacity spectrum, in the parameter range, where the fugacity spectrum obeys a gap between the two principally unconnected spectra close above the critical temperature, which leads to spontaneous symmetry breaking. Thus, assuming Boltzmann equilibrium turns out to be a necessary and sufficient condition for spontaneous symmetry breaking of the quantum field \cite{ref-10} starting from temperatures approaching the critical temperature for Bose-Einstein condensation with a finite coherence time 
of counter-propagating wave fields (see Fig. \ref{figthree}). 

The probabilistic modeling of the dilute Bose gas of $N$ particles below the critical temperature is described by the equation of conditional probability 

\begin{equation}
\frac{p( \mu_A)}{p( \mu_B)} = \frac{{\rm e}^{-\beta\vert\mu_A\vert}}{{\rm e}^{-\beta\vert\mu_B\vert}} = \frac{{\rm e}^{t_B}}{{\rm e}^{t_A}} , 
\label{time_transition}
\end{equation}\\
in the present numerical framework, which defines the relative probability of a quantum particle to switch from a state A with corresponding chemical potential $\mu_A$ to a state B with chemical potential $\mu_B$ for a Bose gas at thermal equilibrium, where the chemical potential is defined by the intrinsic equation

\begin{equation}
\Delta N = \sum_{j\ne0}z^j(\mu)\left[\prod_{k = x,y,z}\frac{1}{1-e^{-j\beta\hbar\omega_k}}-1\right] + \mathcal{O}(a).
\label{ptn}
\end{equation}\\
In Eq. (\ref{ptn}), $\Delta N = (N - N_0)$ is the difference between $N$ the total number of particles, and $N_0$ the number of condensate particles. Further, $\mu$ denotes the chemical potential of the Bose gas, and $\omega_k$ the trap frequency in mode direction $k = x, y, z$. For numerical calculations of the conditional probabilities, the single-particle ground state energy (level) $0.5\hbar(\omega_x+\omega_y+\omega_z)$ effectively cancels out in the numerical modeling. 

Equation (\ref{time_transition}) does not necessarily describe a coherent forward or backward propagation in time, but only the probability for a transition (or new event) from a state with (the characteristics) of a quantum state with entropy $t_A$ to a state with entropy $t_B$, that is there is a non-vanishing probability that a transition with $t_A \ge t_B$ is accepted. Most likely, however, a transition is only accepted, if the absolute value $t_B \ge t_A$, which means that the particle tends to populate states with large entropy (Boltzmann equilibrium that defines the forward propagation of time) in the Markov evolution process of the present quantum theory. 
In the present framework, the system is allowed to exchange energy with the environment. i. e. assuming the thermodynamical constraints of the canonical ensemble. 

\section{III. Quantitative analysis}

For quantitative numerical modeling and analysis, the Bose gas is assumed to be dilute and obey a constant number of particles at finite temperature in a harmonic trapping potential to numerically calculate the coherence time (real part of complex time as defined in Eq. (\ref{coherence_time})) of a quantum particle in the Bose gas in SI units, which corresponds to the range of few hundred microseconds at the given parameter space. 
Interactions in the Bose gas are assumed to be negligible in the limit of a nearly ideal gas, where the s-wave scattering length $a$ of the particles in the gas effectively tends to zero. 
From Eq. (\ref{coherence_time}), we learn that the Bose gas then obeys a stable (Boltzmann) equilibrium for absolute values of complex time in the limit, where $\Gamma_1t\rightarrow0$ and $\vert \Gamma_2 t\vert\ll \hbar\beta$ (which corresponds to a small decoherence rate as compared to the upper bound $\hbar\beta$ for equilibration in the Bose gas, where all particles populate the single-particle ground state mode in the ideal gas). 
This indicates that the main decoherence mechanism in the Bose gas is thermalization. 

Applying only the constraint of particle number conservation as defined by the conservation equation in Eq. (\ref{ptn}), numerical sampling of complex time leads to a distribution of typical time scales for $\Gamma_1 t$ (oscillation time) and $\Gamma_2 t$ (coherence time), shown in Fig. \ref{figtwo}. 
From the sampling of complex time, it is observed that $\Gamma_1 t$, proportional to the real part of complex time, is distributed around zero (with a vanishing width of approximately $0.03~\mu s$) in the given parameter range (as defined in Fig. \ref{figtwo} - upper panel). 
The real part $\Gamma_1 t$ defines the average oscillation frequency in the fugacity spectrum of Fig. \ref{figtwo}, which means that the quantum particle (with constant particle number at finite temperature) quickly couples to excited single particle quantum states, 
related to $\Gamma_1 t\sim0\pm0.03 ~\mu s$ - in the form of a standing wave. 
The distribution of the time scale $\Gamma_2 t$, the directional imaginary part of complex time (that is the effective coherence time for the particle weighted with the decay constant $\Gamma_2$), is shown in the lower panel of Fig. \ref{figtwo}. 
The typical range of the time scale $\Gamma_2 t$ is between $300 ~\mu s$ and $450 ~\mu s$ (see Fig. \ref{figthree}). 
Comparing the distribution of the time scale $\Gamma_2 t$ to the fugacity spectrum in Fig. \ref{figone}, and to the time scale $\Gamma_2 t$ (lower figure), this shows that the coherent coupling of the single particle quantum states decay on a time scale of about $\Gamma_2 t=(300 - 450)~\mu s$. 
To estimate the time scale in a physical framework, it is assumed that the unit time scale of the defined model is defined by $\tau=2\pi\times(\omega_x\omega_y\omega_z)^{-1/3}$, since any frequency of the equilibrated system below that value is effectively zero in the thermally stable quantum system, because of the finite energy uncertainty.

To study the relaxation to thermal equilibrium (in the lower panel of Fig. \ref{figtwo}), the conditional probability condition in Eq. (\ref{time_transition}) is imposed additionally to the constraint of particle number conservation in Eq. (\ref{ptn}) and the condition of constant temperature in the Monte Carlo sampling algorithm for the phases $\Gamma_1 t$ and $\Gamma_2 t$ in Eq. (\ref{coherence_time_2}). 
From these two constraints and the sampling of complex time, one may calculate and analyze the absolute value of complex time, which corresponds to the single-particle entropy (finite quantum coupling to the atomic vapor), related to both the coherence time scale $\Gamma_2 t$ and the oscillation time scale $\Gamma_1 t$, as shown in Fig. \ref{figtwo} for conditional probabilities. 
From Figs. \ref{figtwo}, \ref{figthree} one can extract that the absolute value of the time measure $t$ is distributed around a mean value of approximately $(405\pm15) ~\mu s$ at a given typical temperature of $T=5.0$ nK. 
The distribution of the absolute value of time $t$ has a Gaussian shape and tends to be always distributed around finite quantitative values as compared to no conditional probability (not shown). 
This smoothing is due to the condition that the system has to satisfy the ergodicity assumption of the Boltzmann equilibrium in the present model, where not only the condition of constant temperature and the conservation of particle number, but also the condition of a stable Boltzmann equilibrium is imposed onto the applied sampling method or algorithm, respectively.

\section{IV. Discussion}

In the present theory, it is illustrated that the constraint of particle number conservation naturally leads to the composition of real-valued and purely imaginary valued time to an extended complex-valued number, which defines time in terms of a complex variable to fully capture the system in its full complexity - interacting particles below the critical temperature. 
It is the spectrum of this complex time with the corresponding fugacity spectrum that naturally defines the direction of a composite (complex-valued) time, that is two simultaneous processes described by $\Gamma_1 t$ which describes coherent oscillations of the quantum particle's wave function and $\Gamma_2 t$ which defines the (de-)coherence time of a single quantum particle in the presence of the $N-1$ other particles, in the form of a very tight cone propagating in the positive imaginary direction of complex space on average. 
In that representation, i. e. after coupling a single particle to the rest of the gas within the time scale $\Gamma_2 t\le\hbar\beta$, complex time doesn't obey time-reversal symmetry anymore. 
In the sequel of a quantitative analysis of complex time, it turns out that the real part of weighted complex time is typically distributed around zero (that is almost all particles share the same coherent phase), whereas the imaginary part of weighted complex time mostly entails the system's relaxation properties (finite coherence times, because of spontaneous fluctuations and decoherence). 
In the limit of vanishing temperatures, the process of spontaneous symmetry breaking (characterized by the well-defined relative phase relation between condensate and non-condensate particles) is directly and reproducibly measurable as condensate formation, respectively, as reported for instance in Ref.~\cite{ref-4}. 
Measuring the relative phase between possible condensate and non-condensate states of a quantum particle in the Bose-Einstein condensate, especially in an intermediate temperature range below zero and the critical temperature, Bose-Einstein condensation in nearly-ideal Bose gases is direct evidence for spontaneous symmetry breaking from quantum fluctuations below the critical temperature. 

\section{V. Conclusion}

In conclusion, using a representation of complex time, the (average) coherence time of a quantum field created by a Bose-Einstein condensate is on the order of a few hundred microseconds in the typical parameter regime of Bose gases at thermal equilibrium in harmonic traps below the critical temperature. 
The phase coherence of the quantum field calculated within the present theoretical model can be understood as the spatially averaged atomic field that shows temporal phase correlations. 
Such phase coherence may e.g. be experimentally obtained from the (pulsed) measurement of reflected laser light of the atomic sample. 
In the presented quantum field theory, the coherence time defines a unit scale for the forward propagation of time from destructive partial wave interference and a decay of the average quantum field, whereas the direction of time is due to the equlibration of the Bose gas to the Boltzmann equilibrium. 

Within the presented model, defining a fixed gauge by the Boltzmann equilibrium is a necessary and sufficient condition for spontaneous symmetry breaking in the Bose gas - while each realization of a Boltzmann equilibrium is stable on the coherence time scale of a few hundred microseconds as presented. 
From the spectrum of the fugacity, it is possible to understand and numerically verify that the process of spontaneous symmetry breaking is induced by energy fluctuations that are induced by the reduction of the atomic density below the critical density for Bose-Einstein condensation, which couples (unconnected) 
symmetric and asymmetric parts of the fugacity spectrum. 
In the classical limit, entropy fluctuations tend to be distributed around a ring in the complex spectrum of the fugacity, approaching configurations of time-reversal symmetry for the quantum wave field. 

Starting from the analysis of this article and the works presented in \cite{ref-8, ref-9}, it is interesting to further apply the formalism of this quantum field theory to better understand the (formal) relation of elementary physical models for time evolution to simple artificial intelligence models like the single layer perceptron. Interesting initial discussions on environment-induced dynamics with Andreas Buchleitner, Dominique Delande and Beno\^{i}t Gr\'{e}maud, as well as useful comments on the manuscript by Malte Tichy from Blue Yonder Inc., are acknowledged. 

\acknowledgments
The author acknowledges the financial support from IU Internationale Hochschule for the (freelancer) lecturer position at the university, which has in particular enabled the formulation and editing of the present theory on the finite phase coherence time of a quantum field created by an ideal Bose gas.

\end{document}